\documentclass[reprint,preprintnumbers,showpacs,nofootinbib,amsmath,amssymb,aps,prd,letterpaper,
]{revtex4-1}
\usepackage{graphicx}
\usepackage{dcolumn}
\usepackage{bm}
\usepackage{color}
\usepackage[active]{srcltx}
\usepackage{amsmath,amsfonts,amssymb,amsthm,amstext,amscd,eucal,srcltx}
\usepackage{epsfig,graphicx,bm}
\usepackage{epstopdf, epsf}
\usepackage{dcolumn}
\usepackage{hyperref}
\everymath{\displaystyle}
\begin{document}
\title{A new perspective on Gauge-flation}
\author{Amir Ghalee}
\affiliation{{Department of Physics, Tafresh  University,
P. O. Box 39518-79611, Tafresh, Iran}}
\begin{abstract}
Recently Maleknejad and Sheikh-Jabbari have proposed a new model for inflation with non-Abelian gauge fields (Gauge-flation), and they have studied the model by numerical methods\cite{Sheikh}. In this model, the
isotropy of space-time is recovered by suitable combination of gauge configurations, and a scalar field is constructed  by gauge field and the scale factor, which produces
inflation period. In this work, exact solutions for the scalar field and the Hubble parameter are presented and we provide analytic solutions for the numerical results. We explicitly present Hubble parameter and fields as functions of time and it is also demonstrated that in some conditions they are damped oscillator.
Moreover, reheating period in the model is discussed.
\end{abstract}
\maketitle
\section{\label{sec:level1}INTRODUCTION}
Observations of the cosmic microwave background and large scale structure
are consistent with the isotropic symmetry of space-time and inflation \cite{Weinberg}, a period
of accelerated expansion occurring in early universe. To save isotropic
symmetry of background space-time, many models use a single or multi-scalar
field(s)(inflaton) with \emph{ansatz} that the inflaton field rolls slowly down
its potential (inflationary period) \cite{Weinberg}. Eventually the field(s) oscillates around
the minimum of its potential and decays into light particles(reheating period).
It has been assumed that reheating period took place just after inflationary
period and the field behaved like dust matter in the reheating period \cite{Weinberg}.\\
We have had few models that can be solved exactly in the cosmological context. The well-known example, that can be solved exactly,
is the exponential potential \cite{Weinberg, Ratra} which, with suitable parameters, gives us inflationary solution(power law inflation). Usually, solving field equations are limited to either the inflationary period, with slow roll
approximations, or reheating period.\\
The success of gauge symmetry in the standard particle physics shows that the gauge symmetry
is the correct tool to construct an effective field theory. But, at first glance, it seems
impossible to reconcile the isotropic symmetry with the vectorial nature of gauge fields. However,
recently a new model for inflation, Gauge-flation, based on gauge field
, $A_{~\mu}^a$, where $a,b,\ldots$ and $ \mu, \nu, \ldots$ are used for the indices
of gauge algebra and the space-time respectively, has been proposed by Maleknejad and Sheikh-Jabbari(MS) \cite{Sheikh}. The rotational symmetry in 3d space is
retained by introducing three gauge fields, such that these gauge fields rotate among each other
by $SU(2)$ non-Abelian gauge transformations \cite{Sheikh}. In this model, a scalar field that causes inflation period,
$\psi$, is constructed from the gauge field and the scale factor (see below). MS have studied the model by \textit{numerical} analysis and obtained numerical solutions for the model  \cite{Sheikh}.\\
In this work we investigate the model by \textit{analytic} method. As we will see, the model has some features that allow us to
obtain leading order of fields in all epochs of the universe and not just only in a specific period. We provide an \textit{analytic} expression for MS solutions and we study reheating period in this model. \\
This letter is organized as follows: in \S II we briefly review the model and obtain equations in terms of variables which we
will use in this letter. In \S III we consider a special solution in spectrum of solutions of the model, that the equations can be solved exactly. The solution yields analytic expression for the Hubble parameter and fields for all epochs of the universe. Also we give the general form of solutions for the fields. In \S \textrm{IV} we obtain the analytic form for the MS solutions. Constraints on the solutions from slow roll conditions and observations are discussed in \S \textrm{V}. In \S \textrm{VI}
we discuss about reheating period in the model. We summarize our finding in \S \textrm{VII}.
In Appendix A, we obtain the equation of motion of fields in terms of variables that we use in this letter.
\section{\label{sec:level1}The model}
We work with a general flat-space FRW background
metric with signature $(-+++)$, and reduced Plank unites $8\pi G=1$. Following \cite{Sheikh}, we consider the effective Lagrangian,
\begin{equation}\label{The-Lagrangian}
{\cal L}=\sqrt{-g}\left(-\frac{R}{2}-\frac{1}{4}F^a_{~\mu\nu}F_a^{~\mu\nu}+\frac{\kappa^{2}}{384
}
(\epsilon^{\mu\nu\lambda\sigma}F^a_{~\mu\nu}F^a_{~\lambda\sigma})^2\right),
\end{equation}
where $\epsilon^{\mu\nu\lambda\sigma}$ is the totally antisymmetric
tensor and the strength field $F$ is
\begin{equation}\label{strength}%
F^a_{~\mu\nu}=\partial_\mu A^a_{~\nu}-\partial_\nu
A^a_{~\mu}-g\epsilon^a_{~bc}A^b_{~\mu}A^c_{~\nu}\quad,
\end{equation}
where $\epsilon_{abc}$ is the totally antisymmetric tensor.
Sheikh-Jabbari \cite{Sheikh2} has shown how $F^{4}$
term in \eqref{The-Lagrangian} is obtained
by integrating out a massive axion field in
Chromo$-$Natural inflation \cite{Chromo-Natural Inflation}. Also, it has been argued how other dimension 8 level terms are suppressed \cite{Sheikh2}.\\
To retain isotropy symmetry of
space-time we have to set \cite{Sheikh}
\begin{equation}\label{field}%
A^a_{~\mu}=\left\{
\begin{array}{ll} \phi(t)\delta^a_i\, ,\qquad  &\mu=i
\\
0\,, \qquad &\mu=0\,.
\end{array}\right.
\end{equation}
Applying Einstein's equations with this setup results in \cite{Sheikh}
\begin{equation}\label{Friedmann1}
\begin{split}
\dot{H}&+2H^2=\kappa^{2}\frac{g^2\phi^4\dot{\phi}^2}{a^6} ,\cr
\dot{H}~&=-(\frac{\dot{\phi}^2}{a^2}+\frac{g^2\phi^4}{a^4})\,,
\end{split}
\end{equation}
where $a$ is the scale factor. The equation of motion for $\phi$ is obtained by combination of the equations in\eqref{Friedmann1} \cite{Sheikh}, so it is not an independent equation and we take \eqref{Friedmann1} to study the model . Neither $\phi(t)$ nor $a(t)$ are scalar under general
coordinate transformation, while $\psi(t)=\phi(t)/a(t)$ is a scalar, so we seek solutions
for it. We define a new variable $\alpha(t)$ by writing
\begin{equation}\label{variable-def}
\begin{split}
\frac{\dot{\phi}}{aH}&=\sqrt{\epsilon(t)}\hspace{.05in}\cos\alpha(t) ,\cr
\frac{g\phi^{2}}{a^{2}H}&=\sqrt{\epsilon(t)}\hspace{.05in}\sin\alpha(t),
\end{split}
\end{equation}
where $\epsilon(t)\equiv -\dot{H}/{H^{2}}$ and $\alpha(t)$ is an \emph{arbitrary}
 function of time. Using \eqref{variable-def}, the system of equations in \eqref{Friedmann1} is reduced to
\begin{equation}\label{reduced}
2-\epsilon(t)=\frac{\kappa^{2}}{4}\epsilon^{2}(t)H^{2}\sin^{2}2\alpha(t),
\end{equation}
noting that $0<\epsilon(t)\leq 2$.
Without assumption either on the $H$ or on the $\alpha(t)$ Eq. \eqref{reduced} cannot be solved
to obtain an expression for the Hubble parameter as a function of time(except for $\kappa=0$ that results in $H=1/2t$, "radiation epoch").
To explore the model, we define
\begin{equation}\label{ansatz2}
H\sin2\alpha(t) \equiv\beta f(t) ,
\end{equation}
where $\beta$ is an arbitrary number (dimensionless in Plank units), but $\beta\leq H$.\\
As we will see, the MS solutions can be obtained by a special form for the $f(t)$. Before we study the MS solution
we take a "simple" ansatz for $f(t)$, that not only helps us to understand the MS solutions, but also, it gives information about dynamics of the model.
\section{\label{sec:level1} The simple ansatz}
In this section we use the following simple form for $f(t)$
\begin{equation}\label{ansatz3}
f(t)= \sin\omega t ,
\end{equation}
where $\omega$ is an arbitrary number.
Using Eqs. \eqref{reduced} and \eqref{ansatz3}, we have
\begin{equation}\label{epsilon2}
\epsilon(t)=\frac{\sqrt{1+2k\sin^{2}\omega t}-1}{\frac{k}{2}\sin^{2}\omega t},
\end{equation}
where $k\equiv(\kappa\beta)^{2}$. Note that $\displaystyle \lim_{\omega t\to n\pi}\epsilon(t)=2$, where $n$
is an integer number. $\epsilon(t)$ is a periodic function of time and for $k>4$ it crosses
$\epsilon(t)=1$ line (acceleration expansion phase), and for  $k>8\times10^{4}$ it crosses
$\epsilon(t)=10^{-2}$ line. The $\epsilon(t)$ is designed as in Fig. 1.\\
To understand behaviour of $\epsilon(t)$, we can derive an useful formula for $\epsilon(t)$ in inflationary period(i.e., when $\epsilon(t)\ll1$).
From  Eqs. \eqref{reduced} and \eqref{ansatz2}, $\epsilon(t)$ in this limit becomes
\begin{equation}\label{scaling}
\epsilon(t)\approx\frac{\sqrt{8}}{\sqrt{k}\sin\omega t_{inf}},
\end{equation}
where the subscript "$inf$" denotes that the above expression is valid when $\epsilon(t)\ll1$,
i.e., $0<\omega t_{inf}<\pi$. For large value for $k$, the validity of Eq. \eqref{scaling} is broken when $t$ is very close to $n\pi/\omega$, so the universe almost evolves through acceleration
expansion phases for $k\gg1$, but eventually exits abruptly from inflationary period, as indicated in Fig. 1.\\

\begin{figure}[h]
\includegraphics{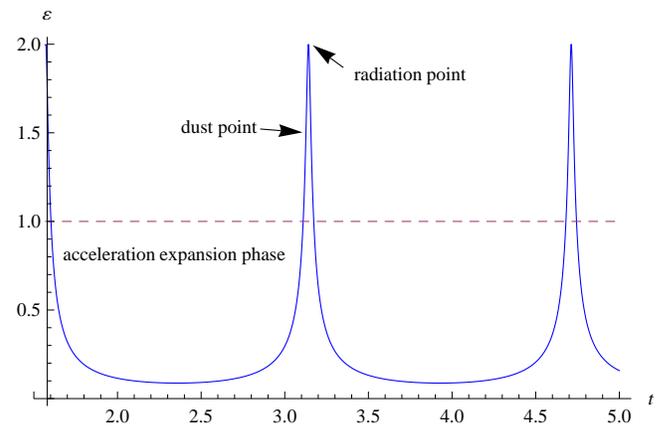}
\caption{\label{fig:epsart} The $\epsilon(t)$ versus $t$ for $k=10^{6}$, $\omega=2$. After inflation,
the universe exits abruptly from
inflationary period.}
\end{figure}

\begin{figure}[h]
\includegraphics{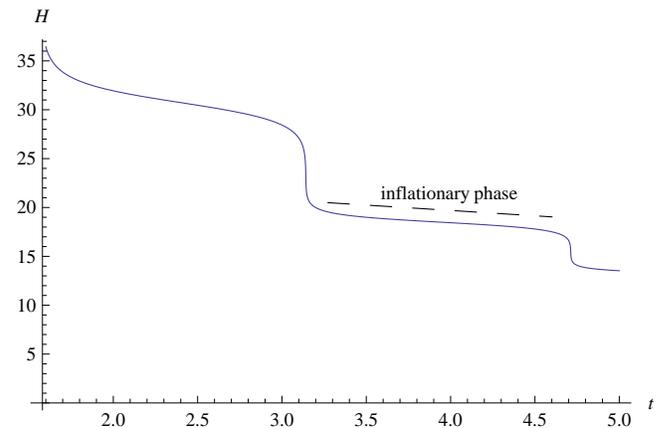}
\caption{\label{fig:epsart}  The Hubble parameter versus $t$ for $k=10^{6}$, $\omega=2$.
Since $-\dot{H}$ increases
with time, the second steep slope is shorter than
the first steep slope.}
\end{figure}

\begin{figure}[h]
\includegraphics{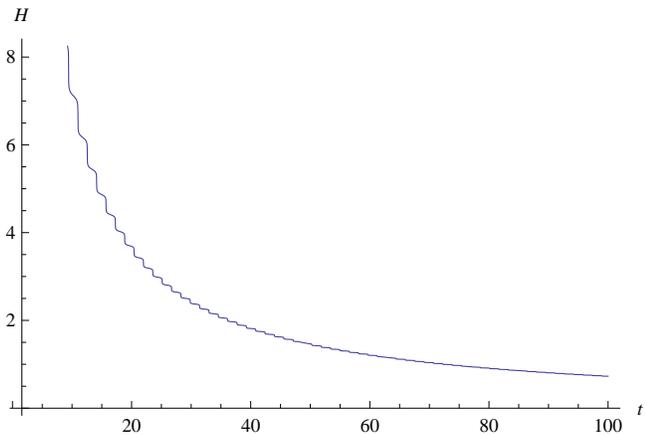}
\caption{\label{fig:epsart} The Hubble parameter versus $t$ for $k=10^{6}$, $\omega=2$.}
\end{figure}

Integration of \eqref{epsilon2} gives
\begin{equation}\begin{split}\label{Hubble}
H(t)={k\omega}\biggr/ 2\biggr(&\left(1-\sqrt{1+k(1-\cos2\omega t)}\right)\cot \omega t\quad +\\
&(1+2k)F[\omega t\,|\!-2k]-E[\omega t\,|\!-2k]\biggr),
\end{split}\end{equation}
where $F[\omega t\,|\!-2k]$ and $E[\omega t\,|\!-2k]$ are
elliptic integrals of the first and
second kind respectively and the specific combination of them in
\eqref{Hubble} is increased monotonically
with time \cite{Abramowitz}.
Also
\begin{equation}\label{Limit}
\displaystyle\lim_{t\to \frac{n\pi}{\omega}}\left((-1+\sqrt{1+k(1-\cos2\omega t)}\right)\cot \omega t=0,
\end{equation}
so the Hubble parameter is damped (but it is singular at $t=0$).
The Hubble parameter is planned as in Fig. 2 and Fig. 3,
where it has gentle slopes when $\epsilon<1$( inflationary phase).
Since $H$ is decreased with time and
$\epsilon(t)$ is a periodic function of time, we conclude that not only $\dot{H}<0$ but also $-\dot{H}$
is decreased with time, as indicated in Fig. 2.\\
If we demand just one inflationary phase for the universe and define $\Delta t_{inf}$ as the duration of
time that inflation takes place, from \eqref{scaling} we have
\begin{equation}\label{constraine}
\omega\Delta t_{inf}<\pi.
\end{equation}
Note that \eqref{constraine} is the upper bound on $\omega\Delta t_{inf}$
that the model predicts itself and it must be consistent with observational constraints (see below).\\
From Eqs.\eqref{variable-def} and \eqref{ansatz2}, we find
\begin{equation}\label{field2}
\psi_{\pm}^{2}(t)=\frac{H\sqrt{\epsilon(t)}}{g}\left(\frac{1\pm\sqrt{1-\frac{\beta^{2}\sin^{2}\omega t}{H^{2}}}}{2}\right)^{\frac{1}{2}},
\end{equation}
where $\epsilon(t)$ and $H$ are given by \eqref{epsilon2} and \eqref{Hubble}
respectively.\\
It is worth to mention that our results in \eqref{epsilon2}, \eqref{Hubble} and \eqref{field2} are valid for all values of $\epsilon(t)$ and this is one of the interesting properties of the model compared with other models for inflation, which use $\epsilon(t)$ as a perturbation parameter and their solutions for the Hubble rate or field(s) are limited to a specific epoch (inflationary period or reheating period). In this sense, the above expressions are exact(non-perturbative) and nonsingular solutions of nonlinear equations \eqref{Friedmann1}.\\
The inverse proportionality of $\psi_{\pm}$ to $\sqrt{g}$, shows that they are not exist in perturbation regime.\\
To understand the behaviour of \eqref{field2} , we use the fact that $\beta\leq H$, and expand \eqref{field2} that gives
\begin{equation}\label{field-approx}
\psi_{+}^{2}(t)=\frac{H\sqrt{\epsilon(t)}}{g} \qquad,\hspace{.3in} \psi_{-}^{2}(t)=\frac{\beta\sqrt{\epsilon(t)}}{2g}\sin\omega t.
\end{equation}
Here the higher terms $\sim \beta^{2}/H^{2}$ have been neglected.
So, $\psi_{+}$ is damped oscillator and $\psi_{-}$ is oscillator (till $\beta \leq H$).
The leading order behaviour of  $\psi^{4}_{\pm}$ are sketched in Fig. 4.\\
Recall that in the reduced Plank units, $H<1$( after Plank time), and $\beta<H$, so $\psi_{-}$
is smaller than $\psi_{+}$ in the inflation period. However, from the Friedman equation, $\rho_{\psi_{\pm}}= 3H^{2}$, we realise that both of them have the same density, and the behaviour of $H$ (in Eq. \eqref{Hubble} and Fig. 3) shows that the cosmic expansion
dilutes the density of fields, although the expectation value of $\psi_{+}$ is greater than $\psi_{-}$
after inflation.\\
For general type of $f(t)$ in \eqref{ansatz2}, $H$ cannot be obtained in terms of well-known functions,
but by algebraic manipulations of equations \eqref{variable-def},\eqref{reduced} and \eqref{ansatz2}, one can show that for leading order
of fields we have
\begin{equation}\label{field-general}
\psi_{+f}^{2}(t)=\frac{H_{f}\sqrt{\epsilon(t)}}{g} \qquad,\hspace{.2in} \psi_{-f}^{2}(t)=\frac{\beta\sqrt{\epsilon(t)}}{2g}f(t),
\end{equation}
where $H_{f}$ is the Hubble rate for $f(t)$.
Since $H_{f}$ is usually unknown, the above expression is formal for $\psi_{+f}$ , but \emph{the leading order of $\psi_{-f}$ is given by \eqref{field-general} for any} \emph{$f(t)$.}\\
Note that Eqs. \eqref{field-general} are valid for all values of $\epsilon(t)$.
The physical meaning of $f(t)$ is clear from \eqref{field-general}, i.e., $f(t)$ shows boundary conditions on the fields.

\begin{figure}[t]
\includegraphics{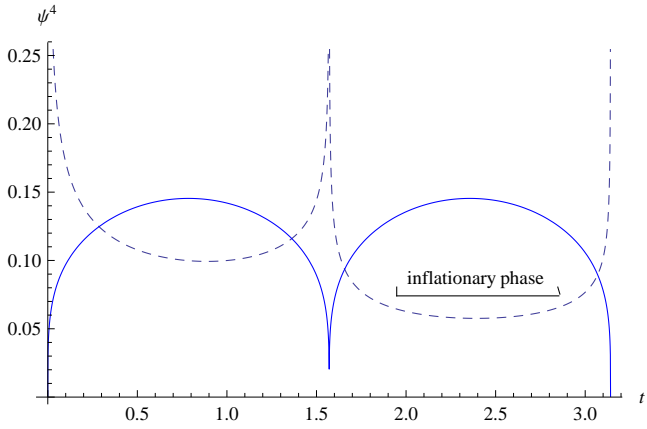}
\caption{\label{fig:epsart} The leading order behaviour of $\psi^{4}_{-}$ (solid line)
and $\psi^{4}_{+}$ (dashed line) for $k=10^{6}$, $\omega=2$, $\beta=g=10^{-6}$}
\end{figure}
\section{\label{sec:level1}The MS solutions}
We pointed out that MS have analysed the model by numerical methods. Here we will show that to rederive the results we must take the following form for $f(t)$
\begin{equation}\label{MS}
f(t)=\left(\frac{\sin(\alpha_{1}t+\alpha_{2}t^{2}+\cdots)}{\alpha_{1}t}\right)^{2}\equiv f_{MS}(t),
\end{equation}
$\epsilon(t)$ and the leading order of $\psi_{-}$ are obtained by \eqref{reduced}, \eqref{ansatz2}, \eqref{field-general} and \eqref{MS} as
\begin{equation}\label{MS epsilon}
\begin{split}
&\epsilon_{MS}(t)=\frac{2\left(\sqrt{1+2kf^{2}_{MS}(t)}-1\right)}{kf^{2}_{MS}(t)} ,\cr
~&\psi_{-MS}^{2}(t)=\frac{\beta\sqrt{\epsilon(t)}}{2g}f_{MS}(t)\,.
\end{split}
\end{equation}
Figures 5 and 6 are obtained by our analytic method and they have the same pattern as MS have obtained\cite{Sheikh}.\\
Note that if $f_{MS}\to 0$ then $\epsilon(t)=2$. Although in this case we cannot obtain an exact expression for the Hubble parameter (for all time), but with \eqref{MS}, we can obtain some information about the MS solution in various regimes.\\
 The most important coefficient of \eqref{MS} in inflationary phase is $\alpha_{1}$. When $\alpha_{1}t<1$ we have $f(t)\approx 1$, so in this regime from \eqref{MS} and \eqref{MS epsilon}, we have
\begin{equation}\label{epsilon MS}
\epsilon_{MS}(t)\approx\frac{2(\sqrt{1+2k}-1)}{k},
\end{equation}
i.e., $\epsilon_{MS}(t)$ is constant in this regime as indicated in Fig. 5. To have $\epsilon(t)\ll10^{-2}$, we must take $k\gg8\times10^{4}$.\\
Therefore the Hubble parameter in this regime is
\begin{equation}\label{hubble MS}
H_{MS}(t)\approx\frac{k}{2(\sqrt{1+2k}-1)t}.
\end{equation}
Hence, from \eqref{field-general} and \eqref{epsilon MS} it follows that
\begin{equation}\label{field-MS}
\begin{split}
\psi_{+MS}^{2}(t)&\approx\sqrt{\frac{k}{2g^{2}(\sqrt{1+2k}-1)}}\hspace{.03in}\frac{1}{t} ,\cr
\psi_{-MS}^{2}(t)~&\approx\frac{\beta}{2g}\sqrt{\frac{2(\sqrt{1+2k}-1)}{k}},,
\end{split}
\end{equation}
so, the leading order of $\psi_{-}$ is constant in this regime, as indicated in Fig. 6.\\
When $\alpha_{1}t\approx\pi$, the numerator in \eqref{MS} equals to zero, and therefore $\epsilon_{MS}\to2$. Just like the simple ansatz in
\S II, for large $k$, the universe almost evolves through inflation period but eventually exits abruptly when $t\to\pi/\alpha_{1}$.\\
Using \eqref{MS}, the Taylor expansion of $\epsilon(t)$ in \eqref{MS epsilon} around $t=\pi/\alpha_{1}$ is
\begin{equation}\label{MS exit}
\epsilon_{MS}(t)\approx2-\frac{\alpha^{4}_{1}k}{\pi^{4}}(t-\frac{\pi}{\alpha_{1}})^{4},
\end{equation}
 The Hubble parameter in this regime can be derived by integration of \eqref{MS exit}, that is
 \begin{equation}\label{ Hubble MS exit}
H_{MS}(t)\approx\frac{1}{2t}\left(1+\frac{\alpha^{4}_{1}k}{10\pi^{4}t}(t-\frac{\pi}{\alpha_{1}})^{5}\right)
\end{equation}
similarly, the leading order of $\psi\pm$ are given by the following formulas
\begin{equation}\label{field-MS-2}
\begin{split}
\psi_{+MS}^{2}(t)&\approx\frac{\sqrt{2}}{2gt}\left(1-\frac{\alpha^{4}_{1}k}{\sqrt{8}\pi^{4}}(t-\frac{\pi}{\alpha_{1}})^{4}\right)\hspace{.03in},\cr
\psi_{-MS}^{2}(t)~&\approx\frac{\beta\sqrt{2}\alpha^{2}_{1}}{2g\pi^{2}}(t-\frac{\pi}{\alpha_{1}})^{2}\,.
\end{split}
\end{equation}
\begin{figure}[h]
\includegraphics{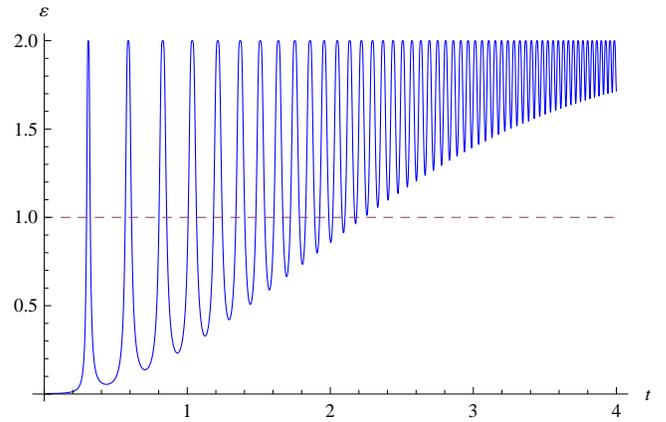}
\caption{\label{fig:epsart} The $\epsilon(t)$ versus $t$ for $k=10^{6}$, $\omega =2$, $\alpha_{1}=10$,$\alpha_{2}=0$,$\alpha_{3}=2$.
Compare with the figures in \cite{Sheikh}}.
\end{figure}
\begin{figure}[b]
\includegraphics{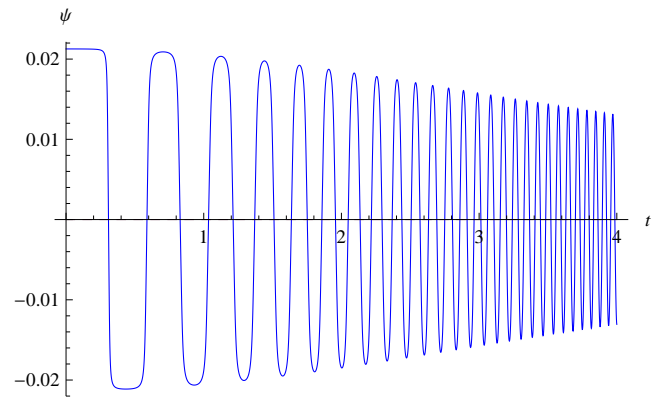}
\caption{\label{fig:epsart} The $\psi_{-}$ versus $t$ with the same parameters as in Fig. 1 and $\beta=g=10^{-6}$. Compare with the figures in \cite{Sheikh}}.
\end{figure}
for $t>1$, the other terms, ($\alpha_{2},\cdots$), in \eqref{MS} are important, and $f_{MS}(t)$ oscillates faster, so $\epsilon(t)$ and $\psi_{\pm}$ oscillate faster.\\
Finally, for $t\gg1$, the denominator in \eqref{MS} is increased, and results in $f_{MS}\approx0$. Therefore in this regime, $\epsilon(t)\to2$, $\psi_{-}\to0$     as indicated in Fig. 5 and Fig. 6. Also, using \eqref{field-general}, we have
\begin{equation}\label{ large t}
H_{MS}(t)\to\frac{1}{2t}\hspace{.03in},\quad\psi_{+}\to\frac{1}{\sqrt{2}g}\frac{1}{t}
\end{equation}

Results from various numerical analysis of the model with various initial values \cite{Sheikh} have similar properties of the MS solutions, that
lead us to the following conjecture about the model:
\textmd{\emph{the MS solution, \eqref{MS}, is the attractor solution of the model.}}
\section{\label{sec:level1} Complementary slow roll
conditions}
If we demand that our fields be inflaton fields, complementary slow roll
conditions are required in inflationary period ,i.e., not only $\epsilon(t)\ll1$ but also
\begin{equation}\label{complementary}
\delta_{\pm f}\equiv-\frac{\dot{\psi}_{\pm f}}{H_{f}\psi_{\pm f}}\ll1 \qquad,\hspace{.3in} \frac{\dot{\delta}_{\pm f}}{H_{f}\delta_{\pm f}}\ll1 .
\end{equation}
For $\psi_{+f}$, the conditions do not have any restriction on the parameters of $\psi_{+f}$, therefore we will focus on $\psi_{-f}$.
\subsection{\label{sec:level1} Complementary slow roll
conditions for the simple ansatz}
For the simple ansatz, \eqref{ansatz3}, the conditions in \eqref{complementary} yield
\begin{equation}\label{slow-roll}
\frac{\omega}{H}\ll2\tan\omega t_{inf} \qquad,\hspace{.3in}  \frac{\omega}{H}\ll\frac{\sin 2\omega t_{inf}}{2}.
\end{equation}
If $0<\omega t_{inf}<{\pi}/{2}$, the conditions in \eqref{slow-roll} are reduced to ${\omega}\ll{H}$, that is agreement with \eqref{constraine}.\\
If we take $\omega\Delta t_{inf}={1}/{2}$, then using the condition ${\omega}\ll{H}$, we obtain
\begin{equation}\label{unce}
H\Delta t_{inf}\gg1.
\end{equation}
The current cosmic microwave background data indicate that
during inflation epoch, $H\lesssim10^{-5}M_{pl}$  \cite{CMB-data}.
If $H\approx 10^{-5}M_{pl}$, the relation \eqref{unce}, implies that $\Delta t_{inf}\gg10^{5}t_{pl}$.\\
Number of $e$-folding is given by
\begin{equation}\label{folding}
N_{e}=\int_{t_{i}}^{t_{f}}Hdt,
\end{equation}
If we use $\omega\Delta t_{inf}={1}/{2}$, then assuming that $t_{f}>10t_{i}$, the numerical integration of the Hubble parameter in \eqref{Hubble}
 shows that to have $N_{e}>60$, we must take $k>4\times10^{6}$.\\
So, if we set $\beta=10^{-6}$ ( in Plank units), then to have sufficient $e$-folds,
we must take $\kappa>2\times10^{9}$.\\
\subsection{\label{sec:level1} Complementary slow roll
conditions for the MS solution}
As for the MS solutions, \eqref{complementary} yields
\begin{equation}\label{slow-roll for MS}
\begin{split}
2(\alpha_{1}t\cot(\alpha_{1}t)-1)&\ll H_{MS}\hspace{.03in} t\hspace{.03in},\cr
\frac{1-(\alpha_{1}t\csc\alpha_{1}t)^{2}}{-1+\alpha_{1}t\cot\alpha_{1}t}~&\ll H_{MS}\hspace{.03in}t\,.
\end{split}
\end{equation}
Here we have neglected other terms that do not have any effect in inflation period.
For $\alpha_{1} t<1$ we have
\begin{equation}\label{function}
\begin{split}
2(\alpha_{1}t\cot(\alpha_{1}t)-1)&\approx -\frac{2(\alpha_{1}t)^{2}}{3}\hspace{.03in},\cr
\frac{1-(\alpha_{1}t\csc\alpha_{1}t)^{2}}{-1+\alpha_{1}t\cot\alpha_{1}t}~&\approx 1+\frac{2}{15}(\alpha_{1}t)^{2})\,.
\end{split}
\end{equation}
Using Eqs. \eqref{hubble MS}, \eqref{slow-roll for MS} and \eqref{function}, the complementary slow roll conditions are reduced to
\begin{equation}\label{compementary MS}
\frac{k}{2(\sqrt{1+2k}-1)}\gg1,
\end{equation}
hence, it is sufficient that $k>8\times10^{4}$.\\
Just like the simple ansatz, if we take $\alpha_{1}\triangle t_{inf}=1/2$, from relations \eqref{slow-roll for MS} and \eqref{function}, we have $H\triangle t_{inf}\gg1$.\\
Number of $e$-folding is given by \eqref{hubble MS} and \eqref{folding} as
 \begin{equation}\label{efolding in ms}
N_{e}\approx\frac{k}{2(\sqrt{1+2k}-1)}\ln\frac{t_{f}}{t_{i}},
\end{equation}
so, for $k>8\times10^{4}$ we have
 \begin{equation}\label{efolding in ms t}
N_{e}>100\ln\frac{t_{f}}{t_{i}},
\end{equation}
The general cosmological perturbation of the model
was developed in \cite{Sheikh}.
\section{\label{sec:level1} Reheating} In the most models for inflation to have
successful reheating period, it is necessary that after inflation period, inflaton(s) behaves like dust matters, and
then decays into relativistic matter. But in the Gauge-flation model the fields can be decayed into relativistic matter, without going to dust matter phase
as we saw in \S IV. One way to see this point is to see the Lagrangian \eqref{The-Lagrangian}, in the inflationary period the $(F\tilde{F})^{2}$ term is dominate, but
after inflation this term is irrelevant, and the second term in \eqref{The-Lagrangian} is dominate after inflation.
Therefore the energy stored
in fields are to be transferred to other fields by thermal
bath of fields.\\
But if we demand that the energy density at
the beginning of radiation epoch
is the same as at the end of inflation, the
thermal bath is not sufficient, due to the expansion of universe.
So, a coupling between fields and matter is needed.\\
We suppose that the fields decay into  relativistic particles, $\chi$,
with decay rate, $\Gamma$, which depends on details of
interactions between the fields with the relativistic particles. Here, we will obtain a bound on $\Gamma$
from conservation of energy. We have \cite{Weinberg}
\begin{equation}\label{energy}
\begin{split}
&\dot{\rho}_{\psi_{\pm}}+3H(\rho_{\psi_{\pm}}+p_{\psi_{\pm}})=-\Gamma\rho_{\psi_{\pm}} ,\cr
&\dot{\rho}_{\chi}+3H(\rho_{\chi}+p_{\chi})=\Gamma\rho_{\psi_{\pm}}.
\end{split}
\end{equation}
Here
\begin{equation}\label{conservation equation}
p_{\chi}/\rho_{\chi}=1/3 ,\hspace{.3in}  p_{\psi_{-}}/\rho_{\psi_{-}}
\equiv w_{eff}=-1+\frac{2}{3}\epsilon(t),
\end{equation}
where $w_{eff}$ is the equation of state. Hence, we can solve \eqref{energy} as
\begin{equation}\label{radiation}
\rho_{\chi}=\frac{\rho_{\psi_{\pm}}(t_{r})}{a^{4}(t)}\frac{\Gamma M^{2}}{\omega H^{2}(t_{r})}\int_{\omega t_{r}}^{\omega t}
a^{4}(\frac{\tau}{\omega})H^{2}(\frac{\tau}{\omega})e^{{\frac{\Gamma}{\omega}}(\tau_{r}-\tau)}\,\mathrm{d}\tau ,
\end{equation}
where $t_{r}$ is the time just after the end of inflation and
$M$ is the scale energy of $H(t)$ that we explicitly show.
After short time $H\rightarrow1/2t$,
and $a(t)\rightarrow a_{0}t^{\frac{1}{2}}$.
By assuming that $\Gamma\gg\omega$, the fields almost immediately
decay into $\chi$, i.e. $H(t_{r})\approx M$, so
\begin{equation}\label{CGS to radiation}
\rho_{\chi}\approx\rho_{\psi_{\pm}(t_{r})}\left(\frac{a_{o}(t_{r})}{a(t)}\right)^{4}.
\end{equation}
Therefore, to have successful inflation and reheating with this scenario, we need $H\Delta t_{inf}\gg1$
and $\Gamma\Delta t_{inf}\gg1$, one can set $\Gamma\approx H$. For the standard
scalar field, to produce a successful radiation epoch after reheating period,
we must take $\Gamma\gg H$ \cite{Weinberg}.
\section{\label{sec:level1} Summary}
We have studied the Gauge-flation by analytic methods and we have investigated the simple (but nontrivial) ansatz that shows the main features of
the model. Then, we have derived formulas for leading order of fields in the model. The formulas are valid in all range of history of the early universe.\\
Using the formulas, we have provided analytic solutions for the MS solutions \cite{Sheikh}, and with the analytic solutions, we
studied some features of the MS solutions which cannot be obtained without analytic methods. Then, we obtained constraints from slow roll conditions on the parameters of the solutions.  \\
Moreover, we studied preheating period in the model and obtained a bound on decay rate of fields, that may be useful for future works.
\begin{acknowledgments}
I am grateful for helpful discussions with F. Arash, H. Asgari, M. M. Sheikh-Jabbari and
A. Maleknejad.

\end{acknowledgments}
\appendix
\section{}
In this letter we give some solutions for Eq. \eqref{reduced}, It is necessary to show that they are also solutions of
the equation of motion for fields. To show this point, in this appendix we will obtain the equation of motion for fields in terms of variables that we use in this letter.\\
The equation of motion can be obtained by variation of \eqref{The-Lagrangian} with respect to the fields as \cite{Sheikh}
\begin{equation}\label{eom}
(1+\kappa g^{2}\frac{\phi^{4}}{a^{4}})\frac{\ddot{\phi}}{a}+(1+\kappa \frac{\dot{\phi}^{2}}{a^{2}})2g^{2}\frac{\phi^{3}}{a^{3}}+(1-3\kappa g^{2}\frac{\phi^{4}}{a^{4}})H\frac{\dot{\phi}}{a}=0.
\end{equation}
But, another standard way, that we use here, to obtain \eqref{eom} is to use the Friedman equations. For what we will do, let us review this method.\\ From \eqref{Friedmann1}, we obtain the following
equations
\begin{equation}\label{Friedmann3}
H^2=\frac{1}{2}\left(\frac{\dot{\phi}^2}{a^2}+\frac{g^2\phi^4}{a^4}+\kappa^{2}\frac{g^2\phi^4\dot{\phi}^2}{a^6}\right),
\end{equation}
and
\begin{equation}\label{Friedmann4}
\dot{H}=-(\frac{\dot{\phi}^2}{a^2}+\frac{g^2\phi^4}{a^4}).
\end{equation}
the derivative of \eqref{Friedmann3} with respect to time results in
\begin{equation}\label{Derivative}
2H\dot{H}=\frac{1}{2}\frac{d}{dt}\left(\frac{\dot{\phi}^2}{a^2}+\frac{g^2\phi^4}{a^4}+\kappa^{2}\frac{g^2\phi^4\dot{\phi}^2}{a^6}\right).
\end{equation}
Substituting Eq. \eqref{Friedmann4} into the left hand side of \eqref{Derivative}, with algebraic manipulations, gives us \eqref{eom}.\\
One way to obtain the equation of motion in terms of variables that we use in this letter, is to substitute variables in Eq. \eqref{eom}, but
it is better to rewrite Eqs. \eqref{Friedmann3} and \eqref{Friedmann4} in terms of our variables, and then to derive the equation of motion.\\
For Eq. \eqref{Friedmann3}, we have
\begin{equation}\label{Friedmann5}
H^2=\frac{H^{2}}{2}\left(\epsilon(t)+\frac{\kappa^{2}}{4}\epsilon^{2}(t)H^{2}\sin^{2}2\alpha(t)\right),
\end{equation}
and for Eq. \eqref{Friedmann4}, we have
\begin{equation}\label{Friedmann6}
\dot{H}=-\epsilon(t)H^{2}.
\end{equation}
The derivative of \eqref{Friedmann5} with respect to time, substituting Eq. \eqref{Friedmann6} into the left hand side of the result, is
\begin{equation}\label{Friedmann7}
\begin{split}
-2H^{3}\epsilon(t)&=H\dot{H}\left(\epsilon(t)+\frac{\kappa^{2}}{4}\epsilon^2(t)H^{2}\sin^{2}2\alpha(t)\right) \cr
&+\frac{1}{2}H^{2}\frac{d}{dt}\left(\epsilon(t)+\frac{\kappa^{2}}{4}\epsilon^{2}(t)H^{2}\sin^{2}2\alpha(t)\right).
\end{split}
\end{equation}
We rearrange Eq. \eqref{Friedmann7} as
\begin{equation}\label{Friedmann8}
\begin{split}
0&=-\epsilon(t)H^{3}\left(-2+\epsilon(t)+\frac{\kappa^{2}}{4}\epsilon^2(t)H^{2}\sin^{2}2\alpha(t)\right) \cr
&+\frac{1}{2}H^{2}\frac{d}{dt}\left(\epsilon(t)+\frac{\kappa^{2}}{4}\epsilon^{2}(t)H^{2}\sin^{2}2\alpha(t)\right).
\end{split}
\end{equation}
Eq. \eqref{Friedmann8} is the equation of motion of fields in terms of our variables.\\
If we rewrite Eq. \eqref{reduced} as
\begin{equation}\label{apreduced}
\epsilon(t)+\frac{\kappa^{2}}{4}\epsilon^{2}(t)H^{2}\sin^{2}2\alpha(t)-2=0,
\end{equation}
then, from Eqs. \eqref{Friedmann8} and \eqref{apreduced}, we have
\begin{equation}\label{final}
-\epsilon(t)H\left(L.H.S. \eqref{apreduced}\right)+\frac{1}{2}\frac{d}{dt}\left(L.H.S. \eqref{apreduced}\right)=0.
\end{equation}
Therefore, solutions of Eq. \eqref{apreduced} are the solutions of Eq. \eqref{final}

\bibliography{apssamp}

\end{document}